\newcommand\citep\cite
\documentclass{iau}
\usepackage{graphicx}
\usepackage{natbib}
\usepackage{aas_macros}
\usepackage{upgreek}

\title[Interpretation of IR variability of AGNs ] 
{Interpretation of IR variability of AGNs \\ in the hollow bi-conical dust outflow model}

\author[Victor L. Oknyansky \& C. Martin Gaskell]   
{Victor L. Oknyansky$^{1,2}$
 \and C. Martin Gaskell$^3$}

\affiliation{$^1$Department of Physics, Faculty of Natural Sciences, University of Haifa, Haifa 3498838, Israel\\ email: {\tt victoroknyansky@gmail.com} \\[\affilskip]
$^2$Sternberg Astronomical Institute, M.V. Lomonosov Moscow State University,  119234, Moscow, Universitetsky pr-t, 13, Russia\\[\affilskip]
$^3$Department of Astronomy and Astrophysics, University of California, Santa Cruz, CA 95064 \\email: {\tt mgaskell@ucsc.edu}}

\pubyear{2023}
\volume{378}  
\pagerange{1--4}
\jname{Black hole winds at all scales}
\editors{G. Bruni,K. Fukumura,
M. Diaz-Trigo  \& S. Laha .}
\begin{document}

\maketitle

\begin{abstract}
 We show that, contrary to simple predictions, most AGNs show at best only a small increase of lags with increasing wavelength in the $J$, $H$, $K$, and $L$ bands. We suggest that a possible cause of this near simultaneity from the near-IR to the mid-IR is that the hot dust is in a hollow bi-conical outflow of which we preferentially see the near side.  In the proposed model sublimation or re-creation of dust (with some delay relative luminosity variations) along our line of sight in the hollow cone as the flux varies could be a factor in explaining the AGN changing-look phenomenon (CL). Variations in the dust obscuration can help explain changes in relationship of H${\beta}$ time delay on L$_{}$uv variability.  The relative wavelength independence of IR lags simplifies the use of IR lags for estimating cosmological parameters.
\keywords{galaxies: active - galaxies: nuclei -
galaxies: Seyfert - infrared: optical and IR variability, time delay, dust, cosmology}
\end{abstract}

\firstsection 
\section{Introduction}

Over half a century has passed since the discovery of optical and near-IR variability of AGNs \citep{Fitch1967, Pacholczyk1971, Penston1971}. This was soon followed by the discovery of time delays between optical ($U$-band) and near-IR variations \citep{Penston1971,Penston1974,Lebofsky1980, Clavel1989,Glass1992,Baribaud1992,Oknyansky1993}. The bolometric luminosity of AGNs is dominated by the ``big blue bump", the thermal emission from the inner regions of the accretion disc.  The IR radiation of AGNs is mostly thermal emission of dust heated by UV radiation from the inner regions. The time delay is interpreted as a consequence of the light-travel time due to the different locations and sizes of the emitting regions \citep{Barvainis1987}. The hot dust is located away from the axis of rotation \citep{Keel1980} in what \citet{Antonucci1984} called a ``torus". Although the shape and structure of this region of hot dust is uncertain, it is commonly depicted in cartoons as being like a doughnut.

Spatially-resolved IR observations of a number of AGNs show that the dust clouds predominantly emitting in the mid-IR to far-IR range are not concentrated in the plane of the galaxy or of the accretion  disc, as might be expected, but in the polar direction \citep{Braatz1993, Cameron1993, Bock2000, Honig2012}. Spatial resolution for IR emission regions is only possible for the nearest AGNs and mostly in the mid- and far-IR regions.  The  most common method for studying the unresolved structure of the emission from warm dust emitting in the mid- to near-IR is reverberation mapping using broad-band IR flux data compared with optical/UV fluxes. Such studies have now been carried out for several dozen AGNs.  These reveal a square root dependence of the IR time delay on the luminosity $(L_{uv})$ for a number of AGNs \citep{Oknyansky1999, Oknyansky2001, Koshida2014, Koshida2017,Minesaki2019}.  This potentially gives us a direct way for estimating cosmological parameters  \citep{Oknyansky1999, Yoshii2014}. Such a luminosity--time delay relationship was predicted theoretically by Barvainis (1987) from radiative equilibrium of the dust.  The IR time delays can correspond to the sublimation radius which corresponds to the  maximal dust  temperature -- approximately 1700–2000 K for graphite grains \citep{Huffman1977, Salpeter1977, BaskinLaor2018}. 

In an individual object (for example, NGC~4151) the time delay varies with the mean $L_{UV}$ level.  This is because of the destruction and replenishment of dust as the high-energy flux varies. These variations of the IR time delay indicate a lag in changes in the size of the hot dust region of about a few years behind $L_{UV}$ variations \citep{Oknyansky2008, Kishimoto2013}. Despite significant growth in theoretical and observational studies of AGNs in the IR, our knowledge of the dust, its origin, kinematics, and detailed morphology remains very incomplete \citep[e.g.,][]{Oknyansky15, Czerny2023}.


\begin{figure}[b]
\begin{center}
 \includegraphics[width=3.5 in]{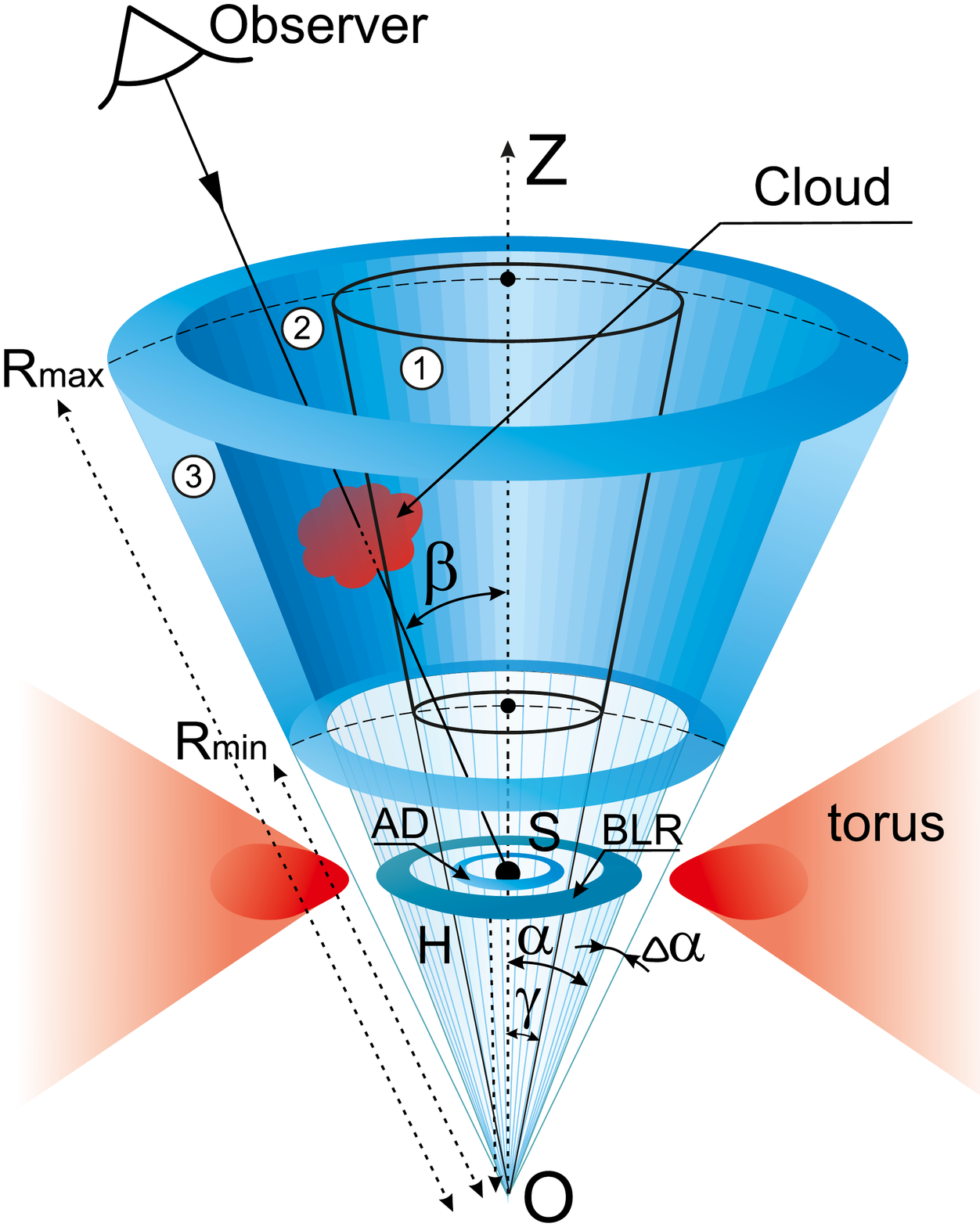} 
 \caption{An idealized schematic showing the near side of the proposed geometry of a hollow bi-conical dust distribution. (The extension of the cones below the equatorial plane is explained in the text.) In this idealized model, the observer is viewing the system at an angle $\upbeta$ to the axis of symmetry. The three zones of dust sublimation or recreation discussed in the text are indicated.  Region number one (half-opening angle $< \upgamma$) indicates where the dust is always mostly sublimated and refractory elements are mostly in the gas phase.  Region number three (half-opening angle $> \upalpha$) corresponds to where the dust will always survive.  The half-opening angles $\upgamma$ and $\upalpha$ defining the boundaries of the intermediate region (number two) depend on the history of flux variations (see text).  In reality, the dust will be in clumps. One such cloud is indicated for illustration purposes and can help explanation the CL events.  This gives partial obscuration of the BLR and the accretion disc (AD). }
  \label{fig1}
\end{center}
\end{figure}


\section{Observed IR time lags}

Observed IR time delays in different  bands have been collected and discussed by \cite{Oknyansky15, Oknyansky2019, Koshida2014, Koshida2017,Minesaki2019} and elsewhere.  From these results, contrary to simple predictions, we find that most AGNs show at best only a small increase of lags in the $J$, $H$, $K$, and $L$ bands with increasing wavelength. Only for 2 objects, GQ Comae and NGC~4151, was a difference of about a factor of three between lags in the $L$ and $K$ bands observed. For NGC~4151 such a big difference was observed only during the  unusually long, very high state which was the only such one during more than a century of monitoring \citep{Oknyansky2013}. The probability of observing an object exactly when it is in its highest state in a century is not large. Statistically it is much more likely that one observes an object in state which is below a very rare high state. Perhaps this is the reason why we see similar IR lags at different wavelengths for most objects. The relative wavelength independence of IR lags simplifies the use of IR lags for estimating cosmological parameters \citep{Oknyansky1999} since it reduces the redshift dependence of the time-lag. 

\section{A hollow, bi-conical, outflow model}

If, as has been widely assumed, the dust clouds are located in the equatorial direction in a flattened, axially-symmetric region (i.e., a ``torus"), and if we are viewing the AGN from a ``type-1" viewing position not too far off-axis, then the IR lag, $\tau$, gives the effective distance, $c\tau$, of the emitting dust clouds from the central source. In this widely-assumed model, similar lags at different IR wavelengths must be interpreted as implying the existence of dust with different temperatures at approximately the same radii. In \cite{Oknyansky15} we propose an alternative explanation, that {\em the dust is in a hollow bi-conical outflow.}  \citep[see  details there,][]{Oknyansky15, oknyansky18}.

In our model the relatively dust-free region would have to have large half-opening angle, $\alpha$ (see Fig.~\ref{fig1}). The structure of IR emission region is axisymmetric due to axial symmetry of the UV radiation given by $F_{uv}(\theta) \sim [cos(\theta) (1 + 2 cos(\theta))/3]^{1/2}$ \citep[see, e.g.,][]{Netzer2015} and also due to dynamic reasons \citep[see, e.g.,][]{Wada2015, Kikubo2020, Czerny2023} the center of the cones (O) is below the equatorial plane. Such geometry can be connected with suspected local origin of the dust in the accretion disk, and also helps with the need to have as much as possible of UV radiation hitting the dust region.   In \cite{Oknyansky15} we showed that IR lags are similar at different radii in the conical outflow and that this explains the relative wavelength independence of the IR time delay. In  Fig~\ref{fig1} we show just the near side of the proposed hollow bi-conical dust distribution.  The corresponding cone on the far side (not shown in Fig. 1) gives a much broader response function.  Also, the accretion disc (AD) in the equatorial plane is optically thick and the inner parts of the far side of the bi-cone will be hidden behind it.  

We consider here what happens in this model when an AGN goes through high and low states. We consider three different general regions of dust sublimation or recreation: 

(1)  A fully-sublimated dust region within half-opening angle $\upgamma$. The dust is sublimated here (matter is in the gas phase) even in low states.  The dust cannot recover quickly in this region.  

(2) A region where the dust is partially sublimated. In this region the dust can survive in some dense parts of clouds where grains are not sublimated to the gas phase. The dust in this region can be recover relatively quickly over a few years when an AGN returns to a low state.

(3) A non-sublimated dust region with a half-opening angle $> \upalpha$ located further out than the sublimation zone corresponding to some past very bright state. 

Based on what has been observed for NGC~4151, very bright states might happen once per decade or even only once in centuries.  If we are lucky and see an object in such an outburst, then we see the sublimation process in this third region.  In these very rarely  observed events there will be significant differences in IR lags for different wavelengths.  As mentioned, we see this for only two objects. This difference in lag can be connected also with different types of dust particles: carbon and silicate. In the bright state, the UV flux can heat the silicate dust to produce enough high re-processing in mid-IR. In this case we see a time delay in the $K$ band corresponding to the {\em carbon} dust sublimation region but the time delay in $L$ band corresponds to the {\em silicate} dust sublimation radius. 

The geometry will certainly not be spherically symmetric. The mostly-dust-free region 1 and the partially-dust-free region 2 will have a form that can be regarded as a sort of three-dimensional figure eight or ``hour-glass" shape such as is seen, for example, in the nebula around the protostar L1527. Depending on the recent luminosity history, the half-opening angle, $\alpha$, of the hollow, dust-free conical region (region 1), as well as, half-opening angle, $\gamma$, of the region where dust is partially sublimated (region 2)  will be larger or smaller. The change in size (i.e., of the half-opening angles, $\alpha$ and $\gamma$) will follow the change in luminosity with a delay of perhaps a few years. In Fig.\ref{fig1} we also show the broad-line region (BLR) which, from the observer's viewpoint, is partially obscured by dust.  

The change in luminosity of the AGN affects things in two ways.  Firstly,  the physical size of the BLR will change -- what is referred to as ``breathing'' (see \citealt{Cackett2006}) -- and secondly, the dust obscuration will change, after some delay of a few years. We see the BLR throuph the dust region and the effective size (depending on dust obscuration) of the BLR has to be following the UV variability with a few years delay.  Similar predictions have been made previously  \citep[see, e.g.,][]{Laor2004,Gaskell2018} and have been reported for NGC~4151 \citep{Chen2023}. The recovery of dust in some clouds in the region (2) can be much faster than the dynamical time and this can help explain CL events in AGNs.

\section{Conclusion}

The lags of IR variability behind UV variability do not show the increase of lag with increase of wavelength predicted by simple models.  We have proposed a solution in which the dust is in a bi-conical outflow.  This not only explains the relative wavelength independence of the IR lags, but the model can be extended to help explain aspects of the AGN ``changing-look" phenomenon and the time dependence of the Balmer line lags from  $L_{uv}$ variations.
 
\section*{Acknowledgements}
 We thank  A.~Cherepashchuk for supporting our research and observations. We are grateful to D.~Chelouche, A.~Laor, and H.~Netzer and for useful discussions. We are thankful to graphic designer Natalia Sinugina for producing Figure 1. This research has been partly supported by Israeli Science Foundation grant no. 2398/19. 
\bibliography{mybib.bib}

\bibliographystyle{apj}
\end{document}